\documentclass[twocolumn,showpacs]{revtex4}
\usepackage{graphicx}
\usepackage{epstopdf}
\begin{document}
\newcommand{\kvec}{\mbox{{\scriptsize {\bf k}}}}
\newcommand{\lvec}{\mbox{{\scriptsize {\bf l}}}}
\newcommand{\qvec}{\mbox{{\scriptsize {\bf q}}}}
\def\eq#1{Eq.\hspace{1mm}(\ref{#1})}
\def\fig#1{Fig.\hspace{1mm}\ref{#1}}
\def\tab#1{Tab.\hspace{1mm}\ref{#1}}
\title{
---------------------------------------------------------------------------------------------------------------\\
Non-BCS temperature dependence of energy gap in thin film electron-doped cuprates 
}
\author{R. Szcz\c{e}{\'s}niak$^{\left(1,2\right)}$}
\email{szczesni@wip.pcz.pl}
\author{A. P. Durajski$^{\left(1\right)}$}
\email{adurajski@wip.pcz.pl} 
\affiliation{$^1$ Institute of Physics, Cz{\c{e}}stochowa University of Technology, Ave. Armii Krajowej 19, 42-200 Cz{\c{e}}stochowa, Poland}
\affiliation{$^2$ Institute of Physics, Jan D{\l}ugosz University in Cz{\c{e}}stochowa, Ave. Armii Krajowej 13/15, 42-200 Cz{\c{e}}stochowa, Poland}
\date{\today}
\begin{abstract}
We investigate the dependence of the energy gap ($G$) on the temperature ($T$) for the electron-doped high-temperature superconductors. 
The following compounds, in the form of the thin films, have been taken into consideration: 
$\rm La_{2-x}Ce_xCuO_{4}$ (LCCO), $\rm Pr_{2-x}Ce_xCuO_{4}$ (PCCO), and $\rm Nd_{2-x}Ce_xCuO_4$ (NCCO). 
It was found that $G\left(T\right)$ deviates from the BCS prediction more, if a concentration of cerium assumes the lower values.
For the lowest concentration (in the case of LCCO and NCCO), the function $G\left(T\right)$ is not quite like the BCS curve, which is connected with the existence of the residual Nernst region. Next, it has been pointed out that the NCCO superconductor becomes structurally unstable for the maximum concentration of cerium, which is leading to the anomalous dependence of the energy gap on the temperature and the induction of the wide Nernst region. 
\end{abstract}
\maketitle

\section{Introduction}

The high-temperature superconductivity in cuprates has been discovered in 1986 \cite{Bednorz1986A}, \cite{Bednorz1988A}. 
Cuprates can be divided into two groups: the hole-doped and the electron-doped superconductors \cite{Wu1987A}, \cite{Maeda1988A}, \cite{Tokura1989A}. 
Both families have distinctly different phase diagrams \cite{Dagotto1994A}. 

In the case of the hole-doped superconductors, the narrow antiferromagnetic region can be observed for the low values of the doping ($\delta$). 
With the increase of $\delta$, the distinct superconducting phase induces for the lower temperatures. The Cooper condensate borders from the high values of $T$ with the Nernst region \cite{Lee2006A}, \cite{Wang2006A}. Above, the pseudogap is observed \cite{Renner1998A}, \cite{Renner1998B}, \cite{Matsuda1999A}, \cite{Hashimoto2007A}. Both the Nernst and the pseudogap region exist at least to the value of $\delta_{\rm max}$, which corresponding to the optimal doping. 
 
On the other hand, the antiferromagnetic phase is very extended in the group of the electron-doped superconductors, and it is bordered with the relatively small region corresponding to the superconducting state \cite{Armitage2010A}. The Nernst region is observed above the superconducting dome \cite{Li2007B}. However, there are no clear results indicating the existence of the pseudogap \cite{Armitage2010A}. 

It should be noted that the hole-doped superconductors characterize the much higher values of the critical temperature ($T_{C}$) than the electron-doped compounds. 
Currently, the highest $T_{C}$ equal to $\sim 160$ K has been measured for ${\rm HgBa_{2}Ca_{2}Cu_{3}O_{8+y}}$, located under the pressure of $\sim 31$ GPa \cite{Gao1994A}. 

In the family of the electron-doped superconductors, the highest critical temperature is equal to $\sim 30$ K, and it has been obtained for  
$\rm La_{2-x}Ce_xCuO_{4}$ (LCCO) \cite{Naito2002A}, \cite{Skinta2003A}. However, ${\rm Sr_{0.9}La_{0.1}CuO_{2}}$ possesses $T_{C}\sim 40$ K, but the synthesize of this compound is extremely difficult \cite{Siegrist1988A}. 

The hole-doped superconductors have the $d$-wave symmetry of the order parameter \cite{Harlingen1995A}, while the symmetry in the electron-doped compounds is still under debate. 
In particular, the results obtained with the help of the penetration depth \cite{Wu1993A}, \cite{Alff1999A}, \cite{Kim2003A}, \cite{Skinta2002B}, the tunneling spectroscopy \cite{Kashiwaya1998A}, \cite{Alff1998A}, and the Raman scattering \cite{Stadlober1995A} point at the $s$-wave symmetry. On the other hand, the other data (the penetration depth \cite{Kokales2000A}, \cite{Prozorov2000A}, ARPES \cite{Armitage2001A}, \cite{Sato2001A}, \cite{Matsui2005A}, the tricrystal experiment \cite{Blumberg2002A}, and the Raman scattering \cite{Cooper1996A}) emphasize the importance of the $d$-wave symmetry. 

In the case of cuprates, it was very quickly realized that the thermodynamic properties of the superconducting phase cannot be accurately calculated in the framework of the BCS theory or the Eliashberg formalism \cite{Bardeen1957A}, \cite{Bardeen1957B}, \cite{Eliashberg1960A}, \cite{Szczesniak2014C}, \cite{Szczesniak2014D}, \cite{Szczesniak2014E}. 
The above fact results from the exotic pairing mechanism and may be explained in two ways. 
The first theory emphasizes the independent role of the strong electron correlations, which are usually modelled by the Hubbard Hamiltonian \cite{Dagotto1994A}, \cite{Hubbard1963A}, \cite{Hubbard1964A}. 
In particular, the performed {\it ab initio} calculations show that in the case of the single copper-oxygen plane the hopping integral between the nearest neighbours ($t_{0}$) is equal to $\sim 400$ meV, and the on-site Coulomb energy ($U_{0}$) has the value of $\sim 5$ eV \cite{Hybertsen1990A}. 
Note that this reasoning naturally explains the existence of the antiferromagnetic phase in cuprates. In particular, for the half-filled electron band, the Hubbard Hamiltonian can be reduced to the Heisenberg operator with the antiferromagnetic hopping integral ($J_{0}\equiv 2t^{2}_{0}/U_{0}$) \cite{Dagotto1994A}, \cite{Spalek2015A}. 

Unfortunately, under this scheme, it has been failed to describe the superconducting phase in the consistent way \cite{Imada1989A}, \cite{Imada1991A}, \cite{Hirsch1993A}. For example, there are great difficulties in explaining the energy gap dependence on the temperature for the lower values of $\delta$ \cite{Szczesniak2014E}, \cite{Szczesniak2012D}, \cite{Szczesniak2014F}, \cite{Szczesniak2014H}, \cite{Szczesniak2015A}. The issue, which is not settled, is also the origin of the pseudogap. 
  
The second path of research is based on the experimental results, which underline the importance of the electron-phonon interaction \cite{Franck1994A}, \cite{Kulic2000A}, \cite{Vedeneev1995A}, \cite{Hofer2000A}, \cite{Damascelli2003A}, \cite{Cuk2005A}, \cite{Gweon2004A}. Note that in the present case, the Coulomb energy should be treated as the source of the depairing correlations. 

In order to substantiate the discussion, let us consider the simplest version of the Hubbard model:
\begin{equation}
\label{r01}
H_{0}=\varepsilon_{0}\sum_{i\sigma}n_{i\sigma}+t_{0}\sum_{ij\sigma}c^{\dagger}_{i\sigma}c_{j\sigma}+U_{0}\sum_{i}n_{i\uparrow}n_{i\downarrow},
\end{equation}
where $c_{i\sigma}$ and $c^{\dagger}_{i\sigma}$ denote the annihilation and creation operators for the electron state. The index $i$ labels the spatial lattice site and $\sigma$ is the spin. Additionally: $n_{i\sigma}\equiv c^{\dagger}_{i\sigma}c_{i\sigma}$. The symbol $\varepsilon_{0}$ represents the reference energy. 

The parameters included in the Hamiltonian (\ref{r01}) should be calculated by using the formulas:
\begin{eqnarray}
\label{r02}
\varepsilon_{0}\equiv\int d^{3}{\bf r}\Phi^{\star}_{i}\left({\bf r}\right)\left[-\frac{1}{2}\nabla^{2}_{{\bf r}}+V\left({\bf r}\right)\right]\Phi_{i}\left({\bf r}\right),\\
t_{0}\equiv\int d^{3}{\bf r}\Phi^{\star}_{i}\left({\bf r}\right)\left[-\frac{1}{2}\nabla^{2}_{{\bf r}}+V\left({\bf r}\right)\right]\Phi_{j}\left({\bf r}\right),\\
U_{0}\equiv\int\int d^{3}{\bf r}_{1}d^{3}{\bf r}_{2}|\Phi_{i}\left({\bf r}_{1}\right)|^{2}V\left({\bf r}_{1}-{\bf r}_{2}\right)|\Phi_{i}\left({\bf r}_{2}\right)|^{2},
\end{eqnarray}
where the standard notation has been used as well as the Born-Oppenheimer approximation \cite{Kolos1998A}. 

The Wannier function $\Phi_{i}\left({\bf r}\right)$ depends on the instantaneous position of the atom. Therefore, having regard to the lattice vibrations, the Hubbard Hamiltonian has to be generalized as follows: $H\rightarrow H_{0}+\delta H$, where: 
\begin{eqnarray}
\label{r03}
\delta H&=&g_{\varepsilon_{0}}\sum_{i\sigma}n_{i\sigma}\phi_{i}+q_{t_{0}}\sum_{ij\sigma}c^{\dagger}_{i\sigma}c_{j\sigma}\phi_{i}\\ \nonumber
&+&g_{U_{0}}\sum_{i}n_{i\uparrow}n_{i\downarrow}\phi_{i}+\omega_{0}\sum_{i}b^{\dagger}_{i}b_{i}.
\end{eqnarray}

The symbols $g_{\varepsilon_{0}}$, $q_{t_{0}}$, and $g_{U_{0}}$ denote the electron-phonon coupling constants. Additionally: 
$\phi_{i}\equiv b^{\dagger}_{i}+b_{i}$, where $b_{i}$ ($b^{\dagger}_{i}$) is the annihilation (creation) operator of the phonon state. The parameter $\omega_{0}$ represents the value of the maximum phonon frequency. 

Next, the Hamiltonian $H$ can be rewritten in the momentum representation. The canonical transformation eliminating the phonon degrees of freedom allows to obtain the fundamental thermodynamic equation \cite{Szczesniak2012D}: 
\begin{equation}
\label{r1}
1=\left(V+\frac{U}{6}|\Delta\left(T\right)|^{2}\right)
\int^{\omega_{0}}_{-\omega_{0}}d\varepsilon\rho\left(\varepsilon\right)
\frac{\tanh\left(\frac{\beta}{2}\sqrt{\varepsilon^{2}+E^{2}}\right)}{2\sqrt{\varepsilon^{2}+E^{2}}},
\end{equation}
where: $E\equiv \left(V+\frac{U}{6}|\Delta\left(T\right)|^{2}\right)|\Delta\left(T\right)|$. The parameters $V$ and $U$ denote the electron-phonon potential (EPh) and the electron-electron-phonon potential (EEPh), respectively: $V\equiv V_{\varepsilon_{0}}+V_{t_{0}}$ and $U\equiv V_{U_{0}}-U_{0}$. 
 
The order parameter is defined by the formula: 
$\Delta\equiv\sum^{\omega_{0}}_{\kvec}\left<c_{-\kvec\downarrow}c_{\kvec-\uparrow}\right>$, while $c_{\kvec\sigma}$ denotes the annihilation operator for the electron state with the momentum ${\bf k}$. The symbol $\left<...\right>$ represents the thermodynamic average \cite{Fetter1971A}, \cite{Gasser1999A}. The summation over ${\bf k}$ should be performed taking under consideration 
the condition: $\varepsilon_{\kvec}\leq|\omega_{0}|$, where $\varepsilon_{\kvec}$ is the electron band energy. 
The inverse temperature has been defined with the help of the expression: $\beta\equiv 1/k_{B}T$, where $k_{B}$ is the Boltzmann constant.

Let us notice that the band energy for the two-dimensional square lattice can be written as: 
$\varepsilon_{\kvec}=-2t_{0}\left[\cos\left(k_{x}\right)+\cos\left(k_{y}\right)\right]$. Thus, the electron density of states has the form: $\rho\left(\varepsilon\right)=b_{1}\ln|{\varepsilon}/{b_{2}}|$, where $b_{1}\equiv-0.04687t_{0}^{-1}$ and $b_{2}\equiv 21.17796t_{0}$ \cite{Szczesniak2001A}, \cite{Szczesniak2002A}, \cite{Szczesniak2003A}, \cite{Markiewicz1997}, \cite{Goicochea1994A}. 

The presented model can explain (at the qualitative level) the essential properties of the high-temperature superconducting state \cite{Szczesniak2012D}. 
Namely, the high critical temperature is related to the existence of the van Hove singularity in the electron density of states. The van Hove singularity is also responsible for the relatively low isotope coefficient. The anomalously high value of the energy gap 
($G\left(0\right)$, where $G\left(0\right)\equiv 2\left(V+\frac{U}{6}|\Delta\left(0\right)|^{2}\right)|\Delta\left(0\right)|$) in respect to the critical temperature, very often observed for $\delta<\delta_{\rm max}$, can be explained by the high ratio $U/V$. It should be noted that in the considered case, the function $G\left(T\right)$ weakly depends on the temperature, which is also in agreement with the experimental data \cite{Kanigel2007A}. 

\begin{figure}
\includegraphics[width=\columnwidth]{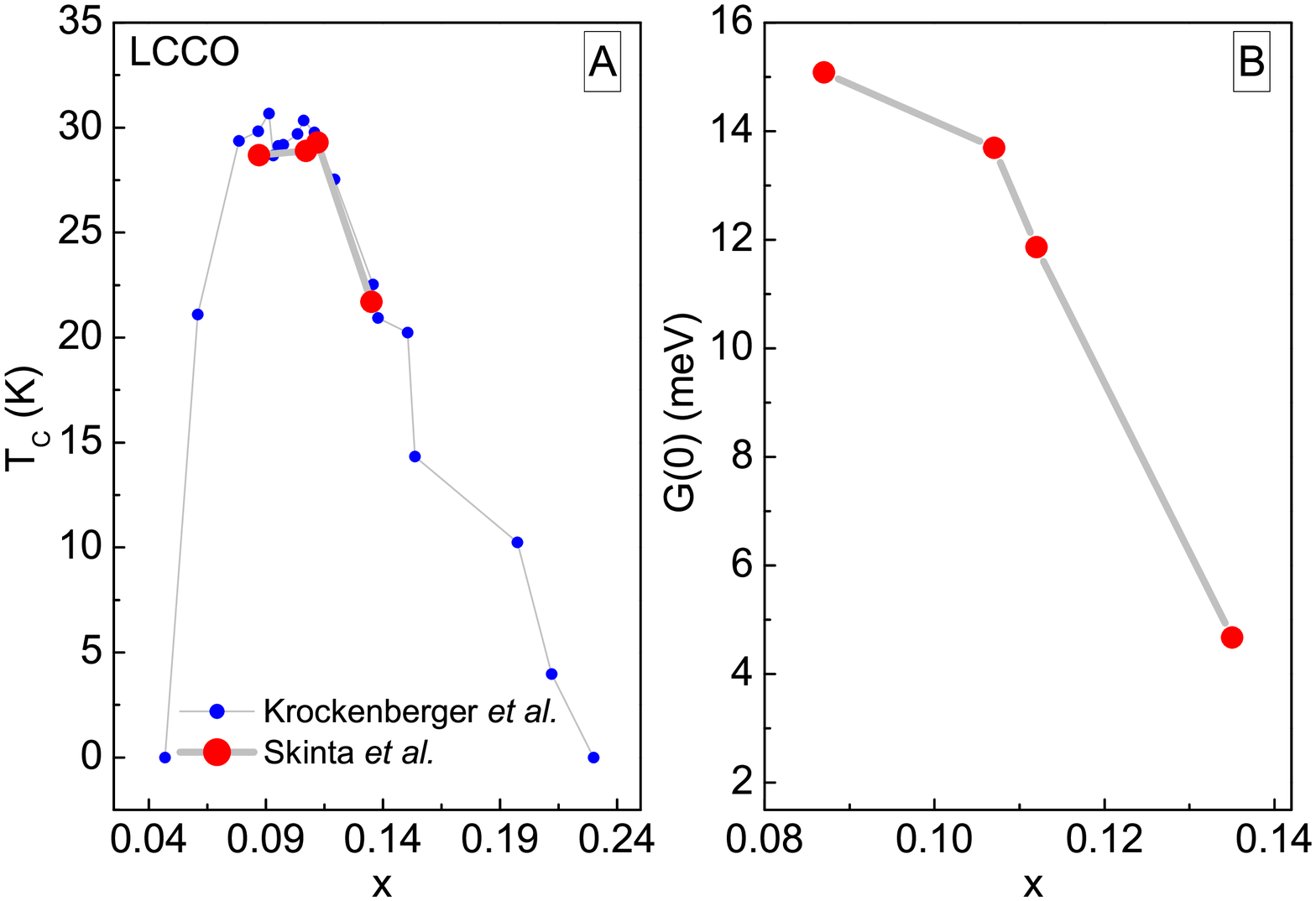} 
\caption{The influence of the cerium concentration on the value of (A) the critical temperature and (B) the energy gap for the superconductor LCCO \cite{Skinta2003A}, \cite{Krockenberger2008A}.}
\label{f1}
\end{figure}
\begin{table}
\caption{\label{t1} The thermodynamic parameters of the LCCO superconductor. Additionally, it has been defined: $U^{\left(0\right)}\equiv\frac{U}{6}|\Delta\left(0\right)|^{2}$.}
\begin{ruledtabular}
\begin{tabular}{c|ccccc}
 x         &$T_{C}$       &G$\left(0\right)$  & $V$                &$U^{\left(0\right)}$     & $T^{\star\star}$ \\
           &      (K)     &   (meV)           & (meV)              &   (meV)                 &     (K)        \\
\hline \\
$0.087$    &$28.7$        &$15.1$             &${\bf 436.6}$       &${\bf 123.9}$            &${\bf 28.9}$\\
$0.107$    &$28.9$        &$13.7$             &${\bf 438.0}$       &${\bf 95.6 }$            &${\bf -}$\\
$0.112$    &$29.3$        &$11.9$             &${\bf 440.7}$       &${\bf 56.7 }$            &${\bf -}$\\
$0.135$    &$21.7$        &$4.7$              &${\bf 387.8}$       &${\bf -50.8}$            &${\bf -}$\\
\end{tabular}
\end{ruledtabular}
\end{table}

The very important feature of the considered approach is related to the existence of the non-zero value of the energy gap above the critical temperature. The performed numerical calculations show that $G\left(T\right)$ vanishes at the Nernst temperature ($T^{\star\star}$) \cite{Szczesniak2012D}. 
 
\vspace*{0.25 cm}

In the present paper, basing on knowledge of the experimental values of the critical temperature and the energy gap close to zero Kelvin, 
the full form of the function $G\left(T\right)$ has been determined. In particular, the following electron-doped superconductors, in the form of thin layers, have been taken into account: LCCO, $\rm Pr_{2-x}Ce_xCuO_{4}$ (PCCO), and $\rm Nd_{2-x}Ce_xCuO_4$ (NCCO). 
Additionally, it has been assumed that $\varepsilon_{0}=0$ and $t_{0}$ is equal to $380$ meV \cite{Andersen1995A}, \cite{Lin2005A}, \cite{Zimmers2007A}, \cite{Hackl2009A}. 

\section{Results}

In \fig{f1} (A)-(B), the dependence of the critical temperature and the low-temperature value of the energy gap ($T=1.6$ K) on the concentration 
of cerium has been presented. The symbols have been prepared on the basis of the experimental data \cite{Skinta2003A}, \cite{Krockenberger2008A} 
(see also \tab{t1}). 

\begin{figure}
\includegraphics[width=\columnwidth]{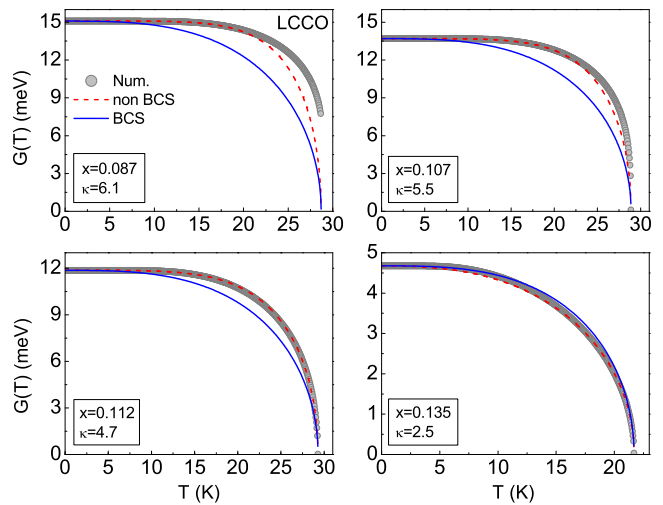} 
\caption{The dependence of the LCCO energy gap on the temperature for the selected values of the Ce concentration.}
\label{f2}
\end{figure}

In the case of the critical temperature, it may be noted that the function $T_{C}\left({\rm x}\right)$ has the typical course observed in the high-temperature superconductors, corresponding approximately to the inverse parabola. The dependence of the energy gap on x is quite different - the value of $G\left(0\right)$ quickly decreases with the increase of the cerium concentration. 
  
Next, based on the data from the paper \cite{Skinta2003A} and with the help of \eq{r1}, the full dependence of the energy gap on the temperature has been calculated. Additionally, it has been assumed that the characteristic phonon frequency is equal to $29$ meV \cite{Gaur2012A}.
The obtained results have been plotted in \fig{f2}. 
It can be seen that for the lowest value for the concentration of cerium, the function $G\left(T\right)$ sharply differs from the BCS dependence of the energy gap on the temperature \cite{Bardeen1957A}, \cite{Bardeen1957B}. In particular, $G$ is much less dependent on the temperature and 
it does not vanish at the critical temperature, but at the Nernst temperature ($T^{\star\star}=28.9$ K). 
For the higher concentrations of cerium, there was no existence of the Nernst area, although the courses of $G$ on $T$ also cannot be reproduced 
in the framework of the BCS model. 

To demonstrate this, we note that the BCS relationship between the energy gap and the temperature can be described by the formula:  

\begin{equation}
\label{r2}
G\left(T\right)=G\left(0\right)\sqrt{1-\left(\frac{T}{T_{C}}\right)^{\kappa}},
\end{equation}
where the exponent $\left[\kappa\right]_{\rm BCS}=3$ \cite{Eschrig2001A}. 

In the case of LCCO, the values of $\kappa$, which the best reproduce the numerical results, have been presented in \fig{f2}. The corresponding curves obtained by using \eq{r2} have been also plotted (the red dashed lines). The blue lines denote the BCS results. 
Additionally, let us pay attention that the values of $\kappa$, with the good approximation, can be calculated from: 
$\kappa\simeq G\left(0\right)/k_{B}T_{C}$. 

\begin{figure}
\includegraphics[width=\columnwidth]{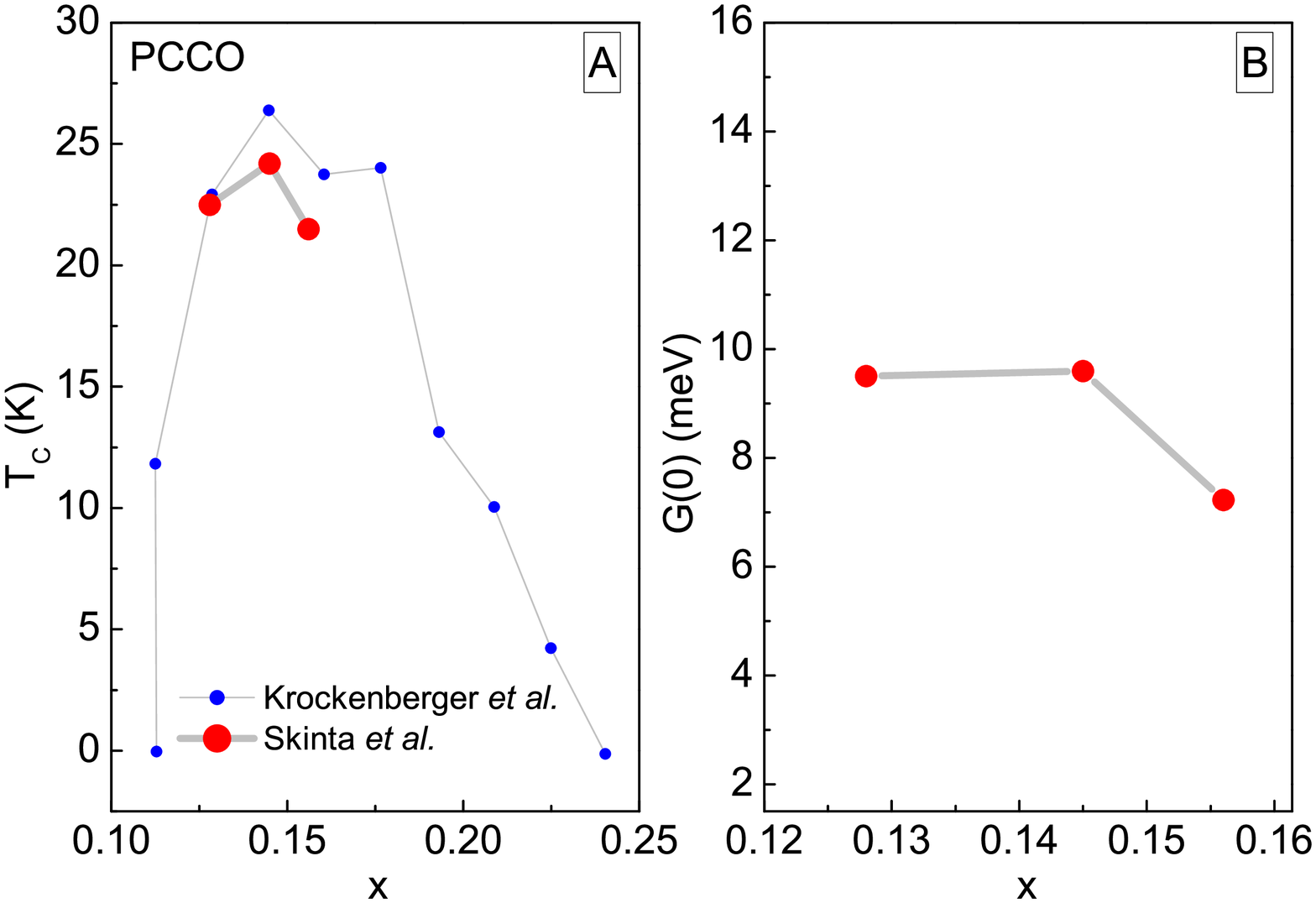} 
\caption{The influence of the cerium concentration on the value of (A) the critical temperature and (B) the energy gap for the PCCO superconductor \cite{Skinta2003A}, \cite{Krockenberger2008A}.}
\label{f3}
\end{figure}
\begin{table}
\caption{\label{t2} The thermodynamic parameters of the PCCO superconductor.}
\begin{ruledtabular}
\begin{tabular}{c|ccccc}
 x            &$T_{C}$    &$G\left(0\right)$            & $V$                 & $U^{\left(0\right)}$   &$T^{\star\star}$   \\
              &   (K)     &   (meV)                     & (meV)               & (meV)                  &      (K)           \\
\hline \\
$0.128$       &$22.5$     &$9.5$                        &$ {\bf 380.4}$       &${\bf     51.5}$        & ${\bf   - }$   \\
$0.145$       &$24.2$     &$9.6$                        &$ {\bf 391.6}$       &${\bf     42.2}$        & ${\bf   - }$   \\
$0.156$       &$21.5$     &$7.2$                        &$ {\bf 373.6}$       &${\bf     12.0}$        & ${\bf   - }$   \\
\end{tabular}
\end{ruledtabular}
\end{table}

Analysing the data plotted in \fig{f2}, it has been found that with the increasing concentration of cerium, the value of  
$\kappa$ falls closer to the result of the BCS model. However, for the sufficiently high concentration (x=0.135), we have: 
$\kappa<\left[\kappa\right]_{\rm BCS}$. This is the interesting case, because it occurs when the effective potential $U^{\left(0\right)}$ changes its sign (see \tab{t1}). From the physical point of view, the obtained result means that electron depairing correlations are stronger than the phonon exchange processes in the EEPh channel ($U_{0}>V_{U_{0}}$). 
 
\vspace*{0.25 cm}

In the next step, the dependence of the energy gap on the temperature for the PCCO superconductor has been determined. 
The study has been based on the experimental results presented in \fig{f3} and in \tab{t2} (data from the paper of Skinta {\it et al.} \cite{Skinta2003A}). During the calculations, it has been assumed that the maximum phonon frequency is equal to $33$ meV \cite{Khlopkin1999A}, \cite{Balci2002A}. 
  
\begin{figure}
\includegraphics[width=\columnwidth]{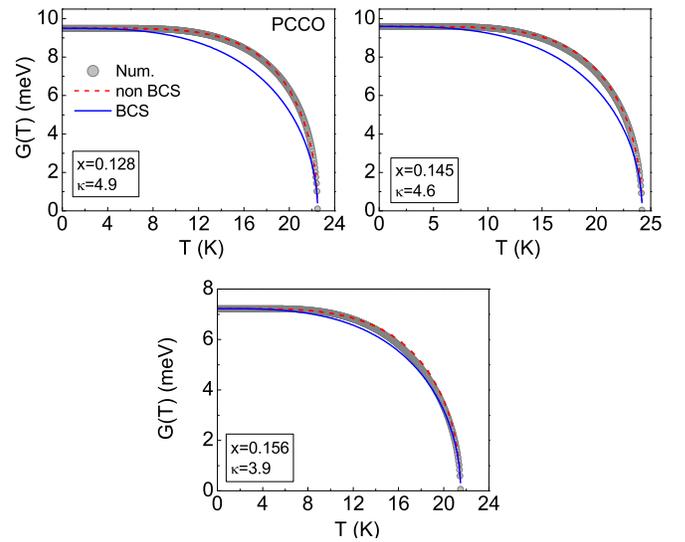} 
\caption{The dependence of the PCCO energy gap on the temperature for the selected values of the Ce concentration.}
\label{f4}
\end{figure}

In the case of the PCCO superconductor, for the low concentrations of cerium, there was no Nernst phase. Additionally, the potential $U^{\left(0\right)}$ does not change the sign for any value of x. As it was in the case for the LCCO superconductor, together with the increasing concentration of cerium, the exponent $\kappa$ decreases and comes closes to the value predicted by the BCS theory (see \fig{f4}).   

\begin{figure}
\includegraphics[width=\columnwidth]{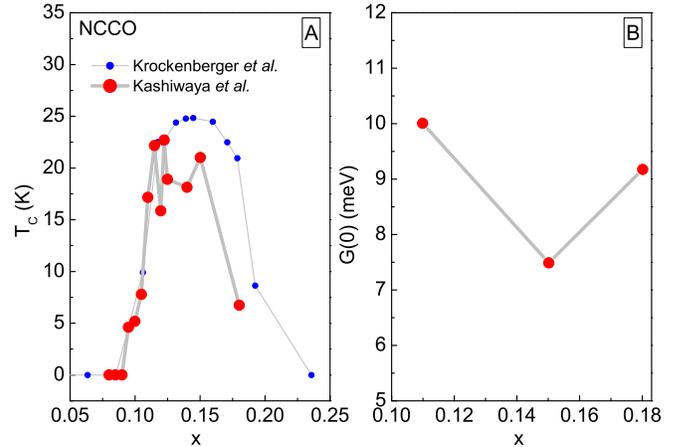} 
\caption{The influence of the cerium concentration on the value of (A) the critical temperature and (B) the energy gap for the NCCO 
superconductor \cite{Krockenberger2008A}, \cite{Kashiwaya2003A}.}
\label{f5}
\end{figure}

\vspace*{0.25 cm}

In the last step, the dependence of the energy gap on the temperature for the NCCO superconductor has been determined. 
The experimental results, on which the calculation has been based, are collected in \fig{f5} and in \tab{t3} (the paper of Kashiwaya {\it et al.} \cite{Kashiwaya2003A}). It has been adopted that the characteristic phonon frequency is equal to $28$ meV \cite{Liu1993A}. 
 
\begin{figure}
\includegraphics[width=\columnwidth]{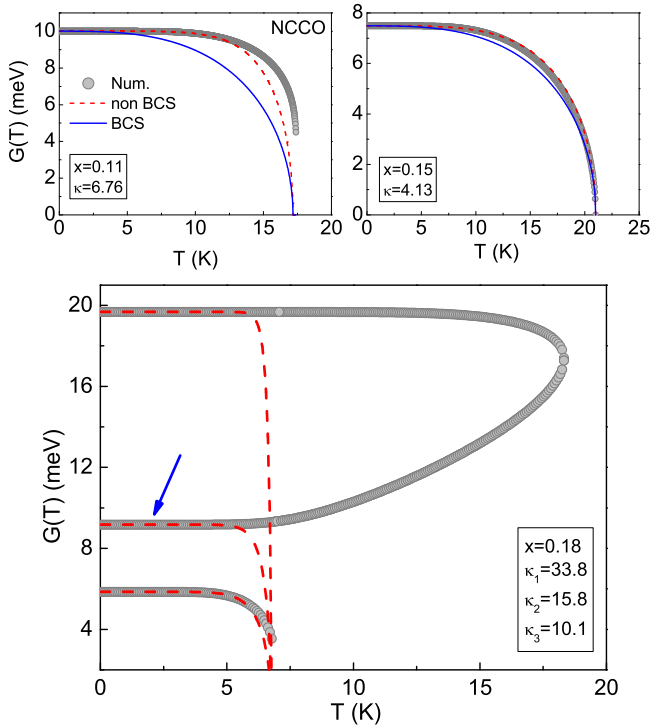} 
\caption{The dependence of the NCCO energy gap on the temperature for the selected values of the Ce concentration. The blue arrow indicates 
the branch of the energy gap, which is realized physically.}
\label{f6}
\end{figure}

The results presented in \fig{f6} prove that for the lowest considered concentration of cerium (x=0.11), above the critical temperature there is the Nernst region, which disappears for $T^{\star\star}=17.4$ K. 

Very unusual results have been obtained for the highest concentration (x=0.18), which follows directly from the experimentally observed 
increase in the low-temperature energy gap value ($T=4.5$ K), (see \fig{f5} (B) or \tab{t3}). 

\begin{table}
\caption{\label{t3} The thermodynamic parameters of the NCCO superconductor.}
\begin{ruledtabular}
\begin{tabular}{c|ccccc}
 x          &$T_{C}$           &$G\left(0\right)$  &$V$                  &$U$                    &$T^{\star\star}$\\
            &     (K)          &   (meV)           & (meV)               & (meV)                 &           (K)  \\
\hline \\
$0.11$      &$17.2$            &$10.0$             &${\bf 356.8}$        &${\bf 107.7}$          &$ {\bf 17.4}$\\
$0.15$      &$21.0$            &$7.5$              &${\bf 386.5}$        &${\bf 22.6 }$          &$ {\bf -}$\\
$0.18$      &$6.8$             &$9.2$              &${\bf 259.0}$        &${\bf 398.5}$          &$ {\bf 18.3}$\\
\end{tabular}
\end{ruledtabular}
\end{table}

In this case, our model predicts that below the critical temperature there are three non-zero branches of $G\left(T\right)$, 
which vary widely in the adopted values (\fig{f6}). Mathematically, it can be shown that the thermodynamic potential minimizes the branch 
with the highest values \cite{Szczesniak2012D}. Experimental data demonstrate, however, that physically realized is the middle branch, which is due to the structural instability of the studied material (above x$>$0.18, it is impossible to synthesize the thin film of NCCO \cite{Kashiwaya2003A}). 

In our opinion, the experimentally observed result is of fundamental importance in the process of falsification of the presented model. 
Namely, the obtained theoretical data clearly predict anomalous dependence of the energy gap on the temperature due to the fact of the increase 
in the value of $G$ together with the increasing $T$. 

In addition, note that above the superconducting state exists the wide Nernst region with the high value of the temperature $T^{\star\star}$ equal to $18.3$ K. 
These predictions can be relatively easy checked experimentally, to which we strongly encourage.  
 
Finally, it should be emphasized that for all concentrations of cerium, the NCCO exponent $\kappa$ significantly differs from the value predicted by the BCS model. 

\section{Summary}

In the paper, the full dependence of the energy gap on the temperature for the high-temperature superconductors LCCO, PCCO, and NCCO  has been determined (the thin films). A wide range of the concentration of cerium has been taken into account. 
 
It has been shown that generally, the curves $G\left(T\right)$ are the more different from the BCS function, than the lower concentration of cerium has been taken into consideration. However, there are the derogation from the above scheme due to the structural instability of the system 
NCCO for ${\rm x}=0.18$. In this case, the increase of the energy gap together with the increasing of the temperature and the high value of 
$T^{\star\star}$ has been predicted. 

For the superconductors LCCO and NCCO (the lowest concentration of cerium) induces the Nernst region, which disappears at the temperature slightly higher than $T_{C}$. 

In the future, the thermodynamic properties of the electron-doped cuprates will be analysed in the framework of the extended Eliashberg approach
\cite{Szczesniak2015B}. The discussed procedure will enable the analysis of the strong-coupling and the retardation effects \cite{Carbotte1990A}. 

\begin{acknowledgments}

The computational calculations have been performed at the Pozna{\'n} Supercomputing and Networking Center. 

\end{acknowledgments}

\bibliography{bibliography}

\end{document}